\documentclass[useAMS,usenatbib]{mn2e}

\title[Accelerating incoherent dedispersion]
	  {Accelerating incoherent dedispersion}
\author[B. R. Barsdell et al.]
  {B. R. Barsdell,$^1$\thanks{Corresponding author:
	  bbarsdel@astro.swin.edu.au}
   M. Bailes,$^1$
   D. G. Barnes,$^1$
   C. J. Fluke$^1$
  \newauthor 
  \\
  $^1$Centre for Astrophysics and Supercomputing, Swinburne University of
  Technology,\\
  PO Box 218, Hawthorn, Australia, 3122}

\pagerange{\pageref{firstpage}--\pageref{lastpage}} \pubyear{2011}

\def\LaTeX{L\kern-.36em\raise.3ex\hbox{a}\kern-.15em
    T\kern-.1667em\lower.7ex\hbox{E}\kern-.125emX}

\usepackage{graphicx}
\usepackage{amsmath}
\usepackage{url}
\usepackage{rotating}
\usepackage{subfigure}

\begin{document}

\label{firstpage}

\maketitle

\begin{abstract}
Incoherent dedispersion is a computationally intensive problem that appears
frequently in pulsar and transient astronomy. For current and future transient
pipelines, dedispersion can dominate the total execution time, meaning its
computational speed acts as a constraint on the quality and quantity of science
results. It is thus critical that the algorithm be able to take advantage of
trends in commodity computing hardware. With this goal in mind, we present
analysis of the `direct', `tree' and `sub-band' dedispersion algorithms with
respect to their potential for efficient execution on modern graphics processing
units (GPUs). We find all three to be excellent candidates, and proceed to
describe implementations in C for CUDA using insight gained from the
analysis. Using recent CPU and GPU hardware, the transition to the GPU provides
a speed-up of 9$\times$ for the direct algorithm when compared to an optimised
quad-core CPU code. For realistic recent survey parameters, these speeds are
high enough that further optimisation is unnecessary to achieve real-time
processing. Where further speed-ups are desirable, we find that the tree and
sub-band algorithms are able to provide 3--7$\times$ better performance at the
cost of certain smearing, memory consumption and development time trade-offs.
We finish with a discussion of the implications of these results for future
transient surveys. Our GPU dedispersion code is publicly available as a C
library at: \url{http://dedisp.googlecode.com/}
\end{abstract}


\section{Introduction}
\label{sec:Introduction}

With the advent of modern telescopes and digital signal processing back-ends,
the time-resolved radio sky has become a rich source of astrophysical
information. Observations of pulsars allow us to probe the nature of neutron
stars (\citealt{LattimerPrakash2004}), stellar populations
(\citealt{BhattacharyaVanDenHeuvel1991}), the galactic environment
(\citealt{GaenslerEtal2008}), plasma physics and gravitational waves
(\citealt{LyneEtal2004}). Of equal significance are transient signals such as
those from rotating radio transients \citep{McLaughlinEtal2006} and potentially
rare one-off events such as `Lorimer bursts'
\citep{LorimerEtal2007,KeaneEtal2011}, which may correspond to previously
unknown phenomena. These observations all depend on the use of significant
computing power to search for signals within long, frequency-resolved time
series.

As radiation from sources such as pulsars propagates to Earth, it interacts with
free electrons in the interstellar medium. This interaction has the effect of
delaying the signal in a frequency-dependent manner -- signals at lower
frequencies are delayed more than those at higher frequencies.  Formally, the
observed time delay, $\Delta t$, between two frequencies $\nu_1$ and $\nu_2$ as
a result of dispersion by the interstellar medium is given by
\begin{align}
\label{eqn:dispersion_delay}
  \Delta t = k_\textrm{DM} \cdot \textrm{DM} \cdot (\nu_1^{-2} - \nu_2^{-2}),
\end{align}
where $k_\textrm{DM} = \frac{e^2}{2\pi m_e c} = 4.148808\times 10^3\;
\textrm{MHz}^2\;\textrm{pc}^{-1}\;\textrm{cm}^3\;\textrm{s}$ is the dispersion
constant\footnote{We note that the dispersion constant is commonly approximated
  in the literature as $1/(2.41\times 10^{-4})\;
  \textrm{MHz}^2\;\textrm{pc}^{-1}\;\textrm{cm}^3\;\textrm{s}$.} and the
frequencies are in MHz. The parameter DM specifies the dispersion measure along
the line of sight in pc cm$^{-3}$, and is defined as
\begin{align}
\label{eqn:dispersion_measure}
  \textrm{DM} \equiv \int_0^d n_e\; \textrm{d}l,
\end{align}
where $n_e$ is the electron number density (cm$^{-3}$) and $d$ is the distance
to the source (pc).

\begin{figure}
\centering
\includegraphics[width=7.9cm]{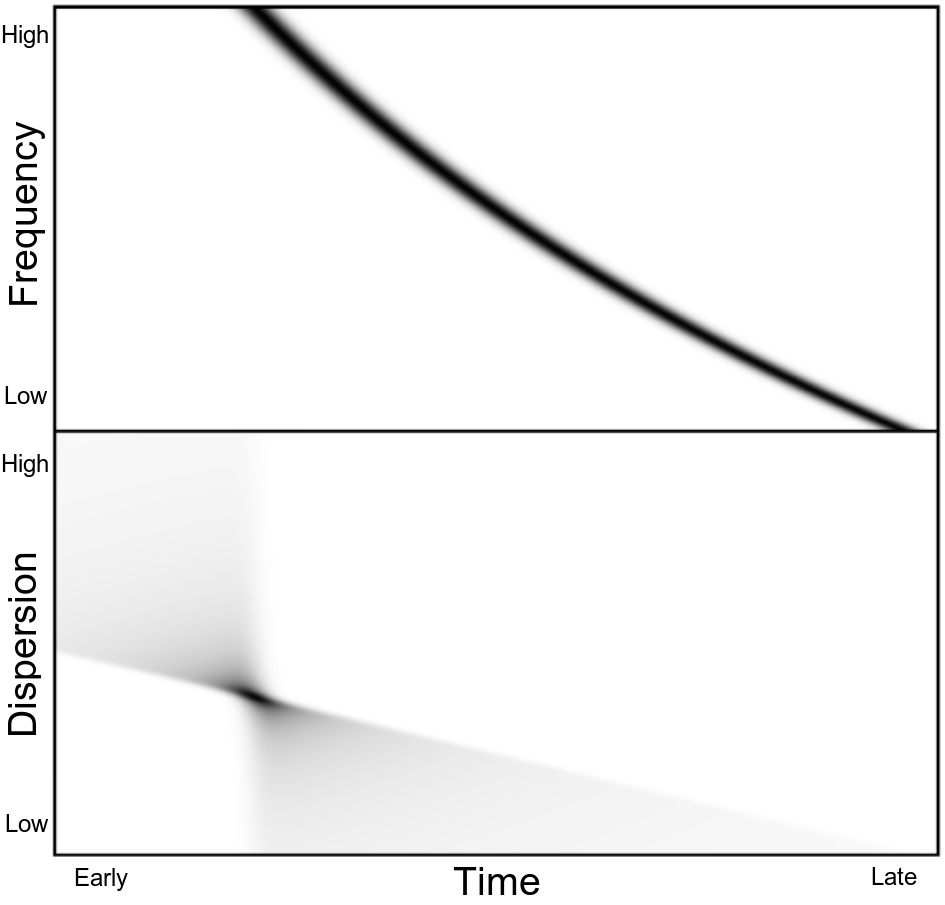}
\caption{An illustration of a dispersion trail (top) and its corresponding
  dedispersion transform (bottom).
  The darkest horizontal slice in the dedispersion transform gives
  the correctly dedispersed time series.
}
\label{fig:Example}
\end{figure}

Once a time-varying source has been detected, its dispersion measure can be
obtained from observations of its phase as a function of frequency; this in
turn allows the distance to the object to be calculated via equation
(\ref{eqn:dispersion_measure}), assuming one has a model for the Galactic
electron density $n_e$. When searching for \textit{new} sources, however, one
does not know the distance to the object. In these cases, the dispersion
measure must be guessed prior to looking for a signal. To avoid excessive
smearing of signals in the time series, and a consequent loss of
signal-to-noise, survey pipelines typically repeat 
the process for many trial dispersion measures. This process is
referred to as a \textit{dedispersion transform}. An example of the
dedispersion transform is shown in Fig.~\ref{fig:Example}.

Computing the dedispersion transform is a computationally expensive task:
a simple approach involves a summation across a band of, e.g., $\sim10^3$
frequency channels for each of $\sim 10^3$ (typically) dispersion measures,
for each time sample. Given modern sampling intervals of $O(64\mu s)$,
computing this in real-time is a challenging task, especially if the process
must be repeated for multiple beams. The prohibitive cost of real-time
dedispersion has traditionally necessitated that pulsar and transient survey
projects use offline processing.

In this paper we consider three ways in which computation of the dedispersion
transform may be accelerated, enabling real-time processing at low cost. First,
in Section~\ref{sec:DirectDedispersionRoot} we demonstrate how modern many-core
computing hardware in the form of graphics processing units [GPUs; see
  \citet{FlukeEtal2011} for an introduction] can provide an order of magnitude
more performance over a multi-core central processing unit (CPU) when
dedispersing `directly'. The use of GPUs for incoherent dedispersion is not an
entirely new idea. \citet{DodsonEtal2010} introduced an implementation of such a
system as part of the CRAFT survey. \citet{MagroEtal2011} described a similar
approach and how it may be used to construct a GPU-based real-time transient
detection pipeline for modest fractional bandwidths, demonstrating that their
GPU dedisperser could out perform a generic code by two orders of magnitude.  In
this work we provide a thorough analysis of both the direct incoherent
dedispersion algorithm itself and the details of its implementation on GPU
hardware.

In Section~\ref{sec:TreeDedispersionRoot} we then consider the use of the `tree'
algorithm, a (theoretically) more efficient means of computing the dedispersion
transform. To our knowledge, this technique has not previously been implemented
on a GPU. We conclude our analysis of dedispersion algorithms in Section
\ref{sec:SubbandDedispersionRoot} with a discussion of the `sub-band' method, a
derivative of the direct method.

In section \ref{sec:Results} we report accuracy and timing benchmarks for the
three algorithms and compare them to our theoretical results. Finally, we
present a discussion of our results, their implications for future pulsar and
transient surveys and a comparison with previous work in Section
\ref{sec:Discussion}.

\section{Direct Dedispersion}
\label{sec:DirectDedispersionRoot}
\subsection{Introduction}
\label{sec:DirectDedispersion}
The direct dedispersion algorithm operates by directly summing frequency
channels along a quadratic dispersion trail for each time sample and
dispersion measure. In detail, the algorithm computes an array of dedispersed
time series $D$ from an input dataset $A$ according to the following equation:
\begin{equation}
\label{eqn:direct}
  D_{d, t} = \sum_\nu^{N_\nu} A_{\nu, t+\Delta t(d,\nu)},
\end{equation}
where the subscripts $d$, $t$ and $\nu$ represent dispersion measure, time
sample and frequency channel respectively, and $N_\nu$ is the total number of
frequency channels. Note that throughout this paper we use the convention that
$\sum_i^N$ means the sum over the range $i=0$ to $i=N-1$. The function $\Delta
t(d,\nu)$ is a discretized version of equation (\ref{eqn:dispersion_delay}) and
gives the time delay relative to the start of the band in whole time samples
for a given dispersion measure and frequency channel:
\begin{align}
\label{eqn:partial_delay}
  \Delta T(\nu) &\equiv \frac{k_\textrm{DM}}{\Delta \tau} \left( \frac{1}{(\nu_0 +
	\nu \Delta\nu)^2} - \frac{1}{\nu_0^2} \right), \\
\label{eqn:delay}
  \Delta t(d,\nu) &\equiv \textrm{round}\left( \textrm{DM}(d) \Delta T(\nu)
    \right),
\end{align}
where $\Delta \tau$ is the time difference in seconds between two
adjacent samples (the \textit{sampling interval}), $\nu_0$ is the frequency in
MHz at the start of the band, $\Delta \nu$ is the frequency difference in MHz
between two adjacent channels and the function round$(x)$ means $x$ rounded to
the nearest integer.
The function or array DM$(d)$
is used to specify the dispersion measures to be computed. Note that the
commonly-used central frequency, $\nu_c$, and bandwidth, BW, parameters are
related by $\textrm{BW}\equiv N_\nu \Delta\nu$ and
$\nu_c\equiv\nu_0+\frac{1}{2}\textrm{BW}$.

After dedispersion, the dedispersed time series $D_{d,t}$ can be searched for
periodic or transient signals.

When dedispersing at large DM, the dispersion of a signal can be such that
it is smeared significantly within a single frequency channel. Specifically,
this occurs when the gradient of a dispersion curve on the time-frequency
grid is less than unity (i.e., beyond the `diagonal'). Once this effect
becomes significant, it becomes somewhat inefficient to continue to dedisperse
at the full native time resolution. One option is to reduce the time
resolution by a factor of two when the DM exceeds the diagonal by adding
adjacent pairs of time samples. This process is then repeated at 2$\times$ the
diagonal, 4$\times$ etc. We refer to this technique as `time-scrunching'. The
use of time-scrunching will reduce the overall computational cost, but can
also slightly reduce the signal-to-noise ratio if the intrinsic pulse width is
comparable to that of the dispersion smear.

\subsection{Algorithm analysis}
\label{sec:DirectAnalysis}
The direct dedispersion algorithm's summation over $N_\nu$ frequency channels
for each of $N_t$ time samples and $N_\textrm{DM}$ dispersion measures gives it
a computational complexity of
\begin{align}
  T_\textrm{direct} = O(N_t N_\nu N_\textrm{DM}).
\end{align}
The algorithm was analysed previously for many-core architectures in
\cite{BarsdellEtal2010}. The key findings were:
\begin{enumerate}
  \item the algorithm is best parallelised over the ``embarassingly parallel''
  dispersion-measure ($d$) and time ($t$) dimensions, with the sum over
  frequency channels ($\nu$) being performed sequentially,
  \item the algorithm has a very high theoretical arithmetic intensity, of the
  same magnitude as the number of dispersion measures computed [typically
  $O(100-1000)$], and
  \item the memory access patterns generally exhibit reasonable locality, but
  their non-trivial nature may make it difficult to achieve a high arithmetic
  intensity.
\end{enumerate}
While overall the algorithm appears well-positioned to take advantage of
massively parallel hardware, we need to perform a deeper analysis to determine
the optimal implementation strategy. The pattern in which memory is accessed
is often critical to performance on massively-parallel architectures, so this
is where we now turn our attention.

While the $d$ dimension involves a potentially non-linear mapping of input
indices to output indices, the $t$ dimension maintains a contiguous mapping
from input to output. This makes the $t$ dimension suitable for efficient
memory access operations via \textit{spatial caching}, where groups of
adjacent parallel threads access memory all at once. This behaviour typically
allows a majority of the available memory bandwidth to be exploited.

The remaining memory access issue is the potential use of \textit{temporal
  caching} to increase the arithmetic intensity of the algorithm. Dedispersion
at similar DMs involves accessing similar regions of input data. By
pre-loading a block of data into a shared cache, many DMs could be computed
before needing to return to main memory for more data. This would increase the
arithmetic intensity by a factor proportional to the size of the shared cache,
potentially providing a significant performance increase, assuming the
algorithm was otherwise limited by available memory bandwidth. The problem
with the direct dedispersion algorithm, however, is its non-linear memory
access pattern in the $d$ dimension. This behaviour makes a caching scheme
difficult to devise, as one must account for threads at different DMs needing
to access data at delayed times. Whether temporal caching can truly be used
effectively for the direct dedispersion algorithm will depend on details of
the implementation.

\subsection{Implementation Notes}
\label{sec:DirectImplementation}

When discussing GPU implementations throughout this paper, we use the terms
`Fermi' and `pre-Fermi' GPUs to mean GPUs of the NVIDIA Fermi architecture and
those of older architectures respectively. We consider both architectures in
order to study the recent evolution of GPU hardware and gain insight into the
future direction of the technology.

We implemented the direct dedispersion algorithm using the C for CUDA
platform\footnote{\url{http://developer.nvidia.com/object/gpucomputing.html}}.
As suggested by the analysis in Section~\ref{sec:DirectAnalysis}, the
algorithm was parallelised over the dispersion-measure and time dimensions,
with each thread summing all $N_\nu$ channels sequentially. During the
analysis it was also noted that the algorithm's memory access pattern exhibits
good spatial locality in the time dimension, with contiguous output indices
mapping to contiguous input indices. We therefore chose time as the
fastest-changing (i.e., $x$) thread dimension, such that reads from global
memory would always be from contiguous regions with a unit stride, maximising
throughput. The DM dimension was consequently mapped to the second (i.e., $y$)
thread dimension.

While the memory access pattern is always contiguous, it is not always
\textit{aligned}. This is a result of the delays, $\Delta t(d,\nu)$,
introduced in the time dimension. At all non-zero DMs, the majority of memory
accesses will begin at arbitrary offsets with respect to the internal
alignment boundaries of the memory hardware. The consequence of this is that
GPUs that do not have built-in caching support may need to split the memory
requests into many smaller ones, significantly impacting throughput to the
processors. In order to avoid this situation, we made use of the GPU's
\textit{texture memory}, which does support automatic caching. On pre-Fermi
GPU hardware, the use of texture memory resulted in a speed-up of around
$5\times$ compared to using plain device memory, highlighting the importance
of understanding the details of an algorithm's memory access patterns when
using these architectures. With the advent of Fermi-class GPUs, however, the
situation has improved significantly. These devices contain an L1 cache that
provides many of the advantages of using texture memory without having to
explicitly refer to a special memory area. Using texture memory on Fermi-class
GPUs was slightly \textit{slower} than using plain device memory (with L1
cache enabled), as suggested in the CUDA programming guide\footnote{The
  current version of the CUDA programming guide is available for download at:
  \url{http://www.nvidia.com/object/cuda_develop.html}}.

Input data with fewer bits per sample than the machine word size (currently
assumed to be 32 bits) were handled
using bit-shifting and masking operations on the GPU. It was found that a
convenient format for working with the input data was to transpose the input
from time-major order to frequency-major order by whole words, leaving
consecutive frequency channels within each word. For example, for the case of
two samples per word, the data order would be:
$(A_{\nu_1,t_1}A_{\nu_2,t_1})$,$(A_{\nu_1,t_2}A_{\nu_2,t_2})$,$ ...$,
$(A_{\nu_3,t_1}A_{\nu_4,t_1})$,$(A_{\nu_3,t_2}A_{\nu_4,t_2})$,$ ...$, where 
brackets denote data within a machine word. This format means that time delays
are always applied in units of whole words, avoiding the need to deal with
intra-word delays.

The thread decomposition was written
to allow the shape of the block
(i.e., number of DMs or time samples per block) to be tuned. We found that for
a block size of 256 threads, optimal performance on a Fermi GPU was achieved
when this was divided into 8 time samples $\times$ 32 DMs. We interpreted this
result as a cache-related effect, where the block shape determines the spread
of memory locations accessed by a group of neighbouring threads spread across
time-DM space, and the optimum occurs when this spread is minimised. On
pre-Fermi GPUs, changing the block shape was found to have very little impact
on performance.

To minimise redundant computations, the functions $\textrm{DM}(d)$ and $\Delta
T(\nu)$ were pre-computed and stored in look-up tables for the given dispersion
measures and frequency channels respectively. Delays were then computed simply
by retrieving values from the two tables and evaluating equation
(\ref{eqn:delay}), requiring only a single multiplication and a rounding
operation. On pre-Fermi GPUs, the table corresponding to $\Delta T(\nu)$ was
explicitly stored in the GPU's \textit{constant memory} space, which provides
extremely efficient access when all threads read the same value (this is always
the case for our implementation, where frequency channels are traversed
sequentially). On Fermi-generation cards, this explicit use of the constant
memory space is unnecessary -- constant memory caching is used automatically
when the compiler determines it to be possible.

To amortize overheads within the GPU kernel such as index calculations, loop
counters and time-delay computations, we allowed each thread to store and sum
multiple time samples. Processing four samples per thread was found to
significantly reduce the total arithmetic cost without affecting memory
throughput. Increasing this number required more registers per thread (a
finite resource), and led to diminishing returns; we found four to be the
optimal solution for our implementation.

Our implementation was written to support a channel ``kill mask'', which
specifies which frequency channels should be included in the computation and
which should be skipped (e.g., to avoid radio frequency interference present
within them). While our initial approach was to apply this mask as a
conditional statement [e.g., \verb@if( kill_mask[channel] ) { sum += data }@],
it was found that applying the mask arithmetically (e.g.,
\verb@sum += data * kill_mask[channel]@) resulted in better performance. This
is not particularly surprising given the GPU hardware's focus on arithmetic
throughput rather than branching operations.

Finally, we investigated the possibility of using temporal caching, as
discussed in the analysis in Section~\ref{sec:DirectAnalysis}. Unlike most
CPUs, GPUs provide a manually-managed cache (known as \textit{shared memory}
on NVIDIA GPUs). This provides additional power and flexibility at the cost of
programming effort. We used shared memory to stage a rectangular section of
input data (i.e., of time-frequency space) in each thread block. Careful
attention was given to the amount of data cached, with additional time samples
being loaded to allow for differences in
delay across a block. The cost of development was significant, and it remained
unclear whether the caching mechanism could be made robust against a variety
of input parameters. Further, we found that the overall performance of the
code was not significantly altered by the addition of the temporal caching
mechanism.
We concluded that the additional overheads involved in handling the non-linear
memory access patterns (i.e., the mapping of blocks of threads in time-DM
space to memory in time-frequency space) negated the performance benefit of
staging data in the shared cache. We note, however, that cacheing may prove
beneficial when considering only low DMs (e.g., below the diagonal), where
delays vary slowly and memory access patterns remain relatively compact.

In theory it is possible that, via careful re-packing of the input data, one
could exploit the bit-level parallelism available in modern computing
hardware in addition to the thread-level parallelism. For example, for 2-bit
data, packing each 2-bit value into 8-bits would allow four values to be
summed in parallel with a single 32-bit addition instruction. In this case,
$\frac{2^8-1}{2^2-1}=85$ 
additions could be performed before one risked integer overflow. To dedisperse
say 1024 channels, one could first sum blocks of 85 channels and then finish
the summation by re-packing the partial sums into a larger data type. This
would achieve efficient use of the available processing hardware, at the cost
of additional implementation complexity and overheads for re-packing and data
management. We did not use this technique in our GPU dedispersion codes,
although our reference CPU code does exploit this extra parallelism by packing
four 2-bit samples into a 64-bit word before dedispersion.

\section{Tree Dedispersion}
\label{sec:TreeDedispersionRoot}
\subsection{Introduction}
\label{sec:TreeDedispersion}
The tree dedispersion algorithm, first described by \citet{Taylor1974}, attempts
to reduce the complexity of the dedispersion computation from $O(N_t N_\nu
N_\textrm{DM})$ to $O(N_t N_\nu \log{N_\nu})$. This significant speed-up is
obtained by first regularising the problem and then exploiting the regularity to
allow repeated calculations to be shared between different DMs. While
theoretical speed-ups of $O(100)$ are possible, in practice a number of
additional overheads arise when working with real data. These overheads, as well
as its increased complexity, have meant that the tree algorithm is rarely used
in modern search pipelines.  In this work we investigate the tree algorithm in
order to assess its usefulness in the age of many-core processors.

In its most basic form, the tree dedispersion algorithm is used to compute the
following:
\begin{align}
\label{eqn:tree}
  D'_{d',t} = \sum_\nu^{N_\nu} A_{\nu, t + \Delta t'(d',\nu)}, \\
  \Delta t'(d',\nu) = \textrm{round}\left( d'\frac{\nu}{N_\nu-1} \right),
\end{align}
for $d'$ in the range $0 \le d' < N_\nu$. The
regularisation is such
that the delay function $\Delta t'(d,\nu)$ is now a linear function of $\nu$
that ranges from $0$ to exactly $d'$ across the band. The DMs computed by the
tree algorithm are therefore:
\begin{align}
\label{eqn:tree_dms}
  \textrm{DM}(d') = \frac{d'}{\Delta T(N_\nu-1)},
\end{align}
where the function $\Delta T(\nu)$ is that given by equation
(\ref{eqn:partial_delay}).

\begin{figure}
\centering
\includegraphics[width=8.5cm]{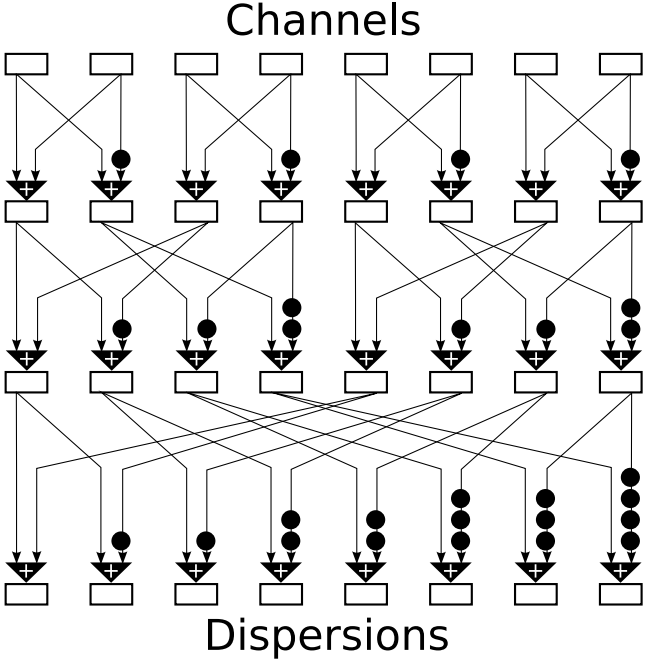}
\caption{Visualisation of the tree dedispersion algorithm. Rectangles
  represent frequency channels, each containing a time series going `into the
  page'. Arrows indicate the flow of data, triangles represent addition
  operations and circles indicate unit time delays into the page.}
\label{fig:TreeAlgorithm}
\end{figure}

The tree algorithm is able to evaluate equation (\ref{eqn:tree}) for $d'$ in the
range $0\le d'<N_\nu$ in just $\log_2{N_\nu}$ steps. It achieves this feat by
using a \textit{divide and conquer} approach in the same way as the well-known
fast Fourier transform (FFT) algorithm. The tree algorithm is visualised in
Fig.~\ref{fig:TreeAlgorithm}.
We define the computation at each step $i$ as follows:
\begin{align}
\label{eqn:tree_algorithm_first}
A^0_{\nu,t} &\equiv A_{\nu,t} \\
A^{i+1}_{2\nu,t}   &= A^i_{\Phi(i,2\nu),t} +
                      A^i_{\Phi(i,2\nu+1),t+\Theta(i,2\nu)} \\
A^{i+1}_{2\nu+1,t} &= A^i_{\Phi(i,2\nu),t} +
                      A^i_{\Phi(i,2\nu+1),t+\Theta(i,2\nu+1)} \\
\label{eqn:tree_algorithm_last}
D'_{d',t} &= D'_{\nu,t} = A^{\log_2{N_\nu}}_{\nu,t}.
\end{align}
The integer function $\Theta(i,\nu)$ gives the time delay for a given
iteration and frequency channel and can be defined as
\begin{align}
\label{eqn:tree_integer_delay}
  \Theta(i,\nu) \equiv [(\nu\:\textrm{mod}\:2^{i+1})+1]/2,
\end{align}
where mod is the modulus operator, and division is taken to be
truncated integer division. The integer function $\Phi(i,\nu)$, which we refer
to as the `shuffle' function, re-orders the indices $\nu$ according to a
pattern defined as follows:
\begin{equation}
\label{eqn:shuffle}
  \Phi(r,\nu) \equiv (\nu\: \textrm{mod}\: 2)\times r +
  \frac{\nu}{2}+\frac{\nu}{2r}\times r,
\end{equation}
where the parameter $r \equiv 2^i$ is known as the \textit{radix}.

While the tree dedispersion algorithm succeeds in dramatically reducing the
computational cost of dedispersion, it has a number of constraints not present
in the direct algorithm:
\begin{enumerate}
  \item the computed dispersion trails are linear in frequency, not quadratic as
    found in nature [see equation (\ref{eqn:dispersion_delay})],
  \item the computed dispersion measures are constrained to those given by
  equation (\ref{eqn:tree_dms}), and
  \item the number of input frequency channels $N_\nu$ (and thus also the number
  of DMs) must be a power of two.
\end{enumerate}
Constraint (iii) is generally not a significant concern, as it is common for
the number of frequency channels to be a power of two, and blank channels can
be added when this is not the case. Constraints (i) and (ii) are more
problematic, as they prevent the computation of accurate and
efficiently-distributed dispersion trails. Fortunately there are ways of
working around these limitations. 

One method is to approximate the dispersion trail with piecewise linear segments
by dividing the input data into sub-bands \citep{ManchesterEtal1996}. Another
approach is to quadratically space the input frequencies by padding with blank
channels as a pre-processing step such that the second order term in the
dispersion trail is removed \citep{ManchesterEtal2001}. These techniques are
described in the next two sections.

\subsubsection{The piecewise linear tree method}
Approximation of the quadratic dispersion curve using piecewise linear segments
involves two stages of computation. If the input data are divided into $N_s$
sub-bands of length
\begin{align}
  N'_\nu = \frac{N_\nu}{N_s},
\end{align}
with the $n^\textrm{th}$ sub-band starting at frequency channel
\begin{align}
  \nu_n &= nN'_\nu,
\end{align}
then from equation (\ref{eqn:tree}) we see that the tree
dedispersion algorithm applied to each sub-band results in the following:
\begin{align}
\label{eqn:subband_result}
  S_{n,d',t} &= \sum_{\nu'}^{N'_{\nu}} A_{\nu_n+\nu',
                                          t+\Delta t'(d',\nu_n+\nu')},
\end{align}
which we refer to as \textit{stage 1} of the piecewise linear tree method.

In each sub-band, we approximate the quadratic dispersion trail with a linear
one. We compute the linear DM in the $n^\textrm{th}$ sub-band that approximates
the true DM indexed by $d$ as follows:
\begin{align}
  d'_n(d) &= \Delta t(d, \nu_{n+1}) - \Delta t(d, \nu_n) \\
\label{eqn:tree_approx_dm_2}
  &= \textrm{round}\left( \textrm{DM}(d)\left[ \Delta T(\nu_{n+1}) - \Delta
    T(\nu_n) \right] \right).
\end{align}
Applying the constraint $d'_n < N'_\nu$ and noting that the greatest dispersion
delay occurs at the end of the band, we obtain a limit on the DM that the basic
piecewise linear tree algorithm can compute. This limit is commonly referred to
as the `diagonal' DM, as it corresponds to a dispersion trail in the
time-frequency grid with a gradient of unity:\footnote{Note that the
  `$\frac{1}{2}$' in equation (\ref{eqn:tree_dm_limit}) arises from the
  round-to-nearest operation in equation (\ref{eqn:tree_approx_dm_2}).}
\begin{align}
\label{eqn:tree_dm_limit}
  \textrm{DM}^\textrm{(piecewise)}_\textrm{diag} =
    \frac{N'_\nu-\frac{1}{2}}{\Delta T(N_\nu) - \Delta T(N_\nu-N'_\nu)}.
\end{align}
A technique for computing larger DMs with the tree algorithm is discussed in
Section~\ref{sec:larger_tree_dms}.

The dedispersed sub-bands can now be combined to approximate the 
result of equation (\ref{eqn:direct}):
\begin{align}
\label{eqn:combining_subbands}
  D_{d,t} &\approx \sum_n^{N_s} S_{n,d'_n(d), t+\Delta t''_n(d)}, \\
  \Delta t''_n(d) &=
  \textrm{round}\left(
    \textrm{DM}(d)
	\sum_m^n \Delta T(\nu_{m+1}) - \Delta T(\nu_m)
  \right).
\end{align}
This forms \textit{stage 2} of the piecewise linear tree computation.

The use of the tree algorithm with sub-banding introduces an additional source
of smearing into the dedispersed time series as a result of approximating the
quadratic dispersion curve with a piecewise linear one. We derive an analytic
upper limit for this smearing in Appendix \ref{sec:TreeError}.

\subsubsection{The frequency-padded tree method}
\label{sec:FrequencyPadding}
An alternative approach to adapting the tree algorithm to quadratic dispersion
curves is to linearise the input data via a change of frequency coordinates.
Formally, the aim is to `stretch'
$\Delta T(\nu)$ [equation (\ref{eqn:partial_delay})] to a linear function
$\Delta T'(\nu') \propto \nu'$. Expanding to first order around $\nu=0$, we
have:
\begin{align}
  \Delta T'(\nu') &= \Delta T(0) +
                   \Delta T(\nu')\left.\frac{d}{d\nu}[\Delta T(\nu)]\right|_0.
\end{align}
The change of variables $\nu \rightarrow \nu'$ is then found by equating
$\Delta T(\nu)$ with its linear approximation, $\Delta T'(\nu')$, and solving
for $\nu'(\nu)$, which gives

\begin{align}
\label{eqn:freq_padding}
  \nu' = \textrm{round}\left( \frac{1}{2}\frac{\nu_0}{\Delta\nu} \left[
             1-\left(1+\frac{\Delta\nu}{\nu_0}\nu \right)^{-2} \right]
         \right).
\end{align}
Evaluating at $\nu = N_\nu$ gives the total number of frequency channels in
the linearised coordinates, which determines the additional computational
overhead introduced by the procedure. Note, however, that this number must be
rounded up to a power of two before the tree dedispersion algorithm can be
applied. For observations with typical effective bandwidths and channel counts
that are already a power of two, the frequency padding technique is unlikely
to require increasing the total number of channels by more than a factor of
two.

In practice, the linearisation procedure is applied by padding the frequency
dimension with blank channels such that the real channels are spaced according
to equation (\ref{eqn:freq_padding}). Once the dispersion trails have been
linearised, the tree algorithm can be applied directly.

The `diagonal' DM when using the frequency padding method corresponds to
\begin{align}
\label{eqn:padded_dm_diag}
\textrm{DM}^\textrm{(padded)}_\textrm{diag} = \frac{1}{\Delta T(1)}.
\end{align}

\subsubsection{Computing larger DMs}
\label{sec:larger_tree_dms}

The basic tree dedispersion algorithm computes exactly the DMs specified by
equation (\ref{eqn:tree_dms}). In practice, however, it is often necessary
to search a much larger range of dispersion measures. Fortunately, there are
techniques by which this can be achieved without having to resort to using the
direct method. The tree algorithm can be made to compute higher DMs by
first transforming the input data and then repeating the dedispersion
computation. Formally, the following sequence of operations can be used to
compute an arbitrary range of DMs:
\begin{enumerate}
  \item Apply the tree algorithm to obtain DMs from zero to
    DM$_\textrm{diag}$.
  \item Impose a time delay across the band.
  \item Apply the tree algorithm to obtain DMs from DM$_\textrm{diag}$ to
    $2$DM$_\textrm{diag}$.
  \item Increment the imposed time delay.
  \item Repeat from step (ii) to obtain DMs up to
    $2^N$DM$_\textrm{diag}$.
\end{enumerate}
The imposed time delay is initially a simple diagonal across the band (i.e.,
$\Delta t=\nu$), and is implemented by incrementing a memory stride value
rather than actually shifting memory. While this method enables dedispersion
up to arbitrarily large DMs, it does not alter the spacing of DM trials, which
remains fixed as per equation (\ref{eqn:tree_dms}).

The `time-scrunching' technique, discussed in Section
\ref{sec:DirectDedispersion} for the direct algorithm, can also be applied to
the tree algorithm. The procedure is as follows:
\begin{enumerate}
  \item Apply the tree algorithm to obtain DMs from zero to DM$_\textrm{diag}$.
  \item Impose a time delay across the band.
  \item Apply the tree algorithm to obtain DMs from DM$_\textrm{diag}$ to
    $2$DM$_\textrm{diag}$.
  \item Compress (`scrunch') time by a factor of 2 by summing adjacent samples.
  \item Impose a time delay across the band.
  \item Apply the tree algorithm to obtain DMs from $2$DM$_\textrm{diag}$ to
	$4$DM$_\textrm{diag}$.
  \item Repeat from step (iv) to obtain DMs up to
	$2^N$DM$_\textrm{diag}$.
\end{enumerate}
As with the direct algorithm, the use of time-scrunching provides a
performance benefit at the cost of
a minor reduction in the signal-to-noise ratio for pulses of intrinsic width
near the dispersion measure smearing time.

\subsection{Algorithm analysis}
\label{sec:TreeAnalysis}

The tree dedispersion algorithm's computational complexity of $O(N_t N_\nu
\log{N_\nu})$ breaks down into $\log_2{N_\nu}$ sequential steps, with each step
involving the computation of $O(N_t N_\nu)$ independent new values, as seen in
equations (\ref{eqn:tree_algorithm_first}) to
(\ref{eqn:tree_algorithm_last}). Following the analysis methodology of
\citet{BarsdellEtal2010}, the algorithm therefore has a \textit{depth
complexity} of $O(\log{N_\nu})$, meaning it contains this many
sequentially-dependent operations. Interestingly, this result matches that of
the direct algorithm, although the tree algorithm requires significantly less
total work. From a theoretical perspective, this implies that the tree
algorithm contains less inherent parallelism than the direct algorithm. In
practice, however, the number of processors will be small relative to the size
of the problem ($N_t N_\nu$), and this reduced inherent parallelism is
unlikely to be a concern for performance except when processing very small
data-sets.

Branching (i.e., conditional statements) within an algorithm can have a
significant effect on performance when targetting GPU-like hardware
\citep{BarsdellEtal2010}. Fortunately, the tree algorithm is inherently
branch-free, with all operations involving only memory accesses and arithmetic
operations. This issue is therefore of no concern in this instance.

The arithmetic intensity of the tree algorithm is determined from the ratio of
arithmetic operations to memory operations. To process $N_t N_\nu$ samples,
the algorithm involves $N_t N_\nu \log_2{N_\nu}$ `delay and add' operations,
and produces $N_t N_\nu$ samples of output. In contrast to the direct
algorithm, where the theoretical arithmetic intensity was proportional to the
number of DMs computed, the tree algorithm requires only $O(\log{N_\nu})$
operations per sample. This suggests that the tree algorithm
may be unable to exploit GPU-like hardware as efficiently as the direct
algorithm. However, the exact arithmetic intensity will depend on constant
factors and additional arithmetic overheads, and will only become apparent
once the algorithm has been implemented. We defer discussion of these results
to Section~\ref{sec:TreeImplementation}.

Achieving peak arithmetic intensity requires reading input data from `slow
memory' into `fast memory' (e.g., from disk into main memory, from main
memory into cache, from host memory into GPU memory etc.) only once, before
performing all computations within fast 
memory and writing the results, again just once, back to slow memory. In the
tree dedispersion algorithm, this means performing all $\log_2{N_\nu}$ steps
entirely within fast memory. The feasibility of this will depend on
implementation details, the discussion of which we defer to Section
\ref{sec:TreeImplementation}. However, it will be useful to assume that some
sub-set of the total computation will fit within this model. We will therefore
continue the analysis of the tree algorithm under the assumption that we are
computing only a (power-of-two) subset, or \textit{block}, of $B_\nu$
channels.

The memory access patterns within the tree algorithm resemble those of the
direct algorithm (see Section~\ref{sec:DirectAnalysis}). Time samples are
always accessed contiguously, with an offset that is essentially arbitrary. In
the frequency dimension, memory is accessed according to the shuffle function
[equation (\ref{eqn:shuffle})] depicted in Fig.~\ref{fig:TreeAlgorithm}, where
at any given step of the algorithm the frequency channels `interact' in pairs,
the interaction involving their addition with different time delays.

With respect to the goal of achieving peak arithmetic intensity, the key issue
for the memory access patterns within the tree algorithm is the extent to
which they remain `compact'. This is important because it determines the
ability to operate on isolated blocks of data independently, which is critical
to prolonging the time between successive trips to slow memory.
In the frequency dimension, the computation of some local 
(power-of-two) sub-set of channels $B_\nu$ involves accessing only other
channels within the same subset. In this sense we can say that the memory
access patterns are `locally compact' in channels. In the time dimension,
however, we note that the algorithm applies compounding delays (equivalent to
offsets in whole time samples). This means that the memory access patterns
`leak' forward, with any local group of time samples always requiring access
to the next group. The amount by which the necessary delays `leak' in time
for each channel is given by the integrated delay in that channel after
$B_\nu$ steps (see Fig.~\ref{fig:TreeAlgorithm}). The total integrated delay
across $B_\nu$ channels is $B_\nu(B_\nu-1)/2$, which is the number of
additional values that must be read into fast memory by the block in order to
compute all $\log_2{B_\nu}$ steps without needing to return to global memory and
apply a global synchronisation.

\subsection{Implementation Notes}
\label{sec:TreeImplementation}

As with the direct algorithm, we implemented the tree algorithm on a GPU in C
for CUDA. For our first attempt, we took a simple approach where each of the
$\log_2{N_\nu}$ steps in the computation was performed by a separate call to a
GPU function (or \textit{kernel}). This approach is not ideal, as it is
preferable to perform more computation on the device before
returning to the host (as per the discussion of arithmetic intensity in
Section~\ref{sec:TreeAnalysis}), but was necessary in order to guarantee
global synchronisation across threads on the GPU between steps. This is a
result of the lack of global synchronisation mechanisms on current GPUs.

Between steps, the integer delay and shuffle functions [equations
(\ref{eqn:tree_integer_delay}) and (\ref{eqn:shuffle})] were evaluated on the
host and stored in look-up tables. These were then copied to constant memory
on the device prior to executing the kernel function to compute the step. The
use of constant memory ensured retrieval of these values would not be a
bottle-neck to performance during the computation of each step of the tree
algorithm.

The problem was divided between threads on the GPU by allocating one thread
for every time sample and every \textit{pair} of frequency channels. This
meant that each thread would compute the delayed sums between two
`interacting' channels according to the pattern depicted in Fig.
\ref{fig:TreeAlgorithm} for the current step.

The tree algorithm's iterative updating behaviour requires that computations
at each step be performed `out-of-place'; i.e., output must be written to a
memory space separate from that of the input to avoid modifying input values
before they have been used. We achieved this effect by using a
double-buffering scheme, where input and output arrays are swapped after each
step.

While the algorithms differ significantly in their details, one point of
consistency between the direct and tree methods is the need to apply time
delays to the input data. Therefore, just as with our implementation of the
direct algorithm, the tree algorithm requires accessing memory
locations that are not aligned with internal memory boundaries. As such, we
took the same approach as before and mapped the input data to the GPU's
texture memory before launching the device kernel.
As noted in Section~\ref{sec:DirectImplementation}, this procedure is
unnecessary on Fermi-generation GPUs, as their built-in caches provide the
same behaviour automatically.

After successfully implementing the tree algorithm on a GPU using a simple
one-step-per-GPU-call approach, we investigated the possibility of computing
\textit{multiple} steps of the algorithm on the GPU before returning to the
CPU for synchronisation. This is possible because current GPUs, while lacking
support for global thread synchronisation, do support synchronisation across
local thread groups (or \textit{blocks}). These thread blocks typically
contain $O(100)$ threads, and provide mechanisms for synchronisation and
data-sharing, both of which are required for a more efficient tree
dedispersion implementation.

As discussed in Section~\ref{sec:TreeAnalysis}, application of the tree
algorithm to a block of $B_\nu$ channels $\times$ $B_t$ time samples
requires caching additional values 
from the next block in time.
We used blocks of
$B_\nu\times B_t=16\times 16$ threads, each loading both their
corresponding data value and required additional values into shared
cache. Once all values have been stored, computation of the $\log_2 B_\nu=4$
steps proceeds entirely within the shared cache. Using larger thread blocks
would allow more steps to be completed within the cache; however, the
choice is constrained by the available volume of shared memory (typically
around 48kB). Once the block
computation is complete, subsequent steps must be computed using the
one-step-per-GPU-call approach described earlier, due to the requirement of
global synchronisations.

While theory suggests that an implementation of the tree algorithm exploiting
shared memory to perform multiple steps in cache would provide a performance
benefit over a simpler implementation, in practice we were unable to achieve a
net gain using this approach. The limitations on block size imposed by the
volume of shared memory, the need to load additional data into cache and the
logarithmic scaling of steps relative to data size significantly reduce the
potential speed-up, and overheads from increased code-complexity quickly
erode what remains. For this reason we reverted to the straight-forward
implementation of the tree algorithm as our final code for testing and
benchmarking.

In addition to the base tree algorithm, we also implemented the sub-band
method so as to allow the computation of arbitrary dispersion measures. This
was achieved by dividing the computation into two stages. In stage 1, the
first $\log_2 N'_\nu$ steps of the tree algorithm are applied to the input
data, which produces the desired $N_\nu/N'_\nu$ tree-dedispersed sub-bands.
Stage 2 then involves applying an algorithm to combine the dedispersed time
series in different sub-bands into approximated quadratic dispersion curves
according to equation (\ref{eqn:combining_subbands}). Stage 2 was implemented
on the GPU in much the same way as the direct algorithm, with input data
mapped to texture memory (on pre-Fermi GPUs) and delays stored in look-up
tables in constant device memory.

The frequency padding approach described in Section~\ref{sec:FrequencyPadding}
was implemented by constructing an array large enough to hold the stretched
frequency coordinates, initialising its elements to zero, and then copying (or
\textit{scattering}) the input data into this array according to equation
(\ref{eqn:freq_padding}). The results of this procedure were then fed to the
basic tree dedispersion code to produce the final set of dedispersed time
series.

Because the tree algorithm involves sequentially updating the entire data-set,
the data must remain in their final format for the duration of the
computation. This means that low bit-rate data, e.g., 2-bit, must be unpacked
(in a pre-processing step) into a format that will not overflow during
accumulation. This is in contrast to the direct algorithm, where each sum is
independent, and can be stored locally to each thread.

\section{Sub-band dedispersion}
\label{sec:SubbandDedispersionRoot}
\subsection{Introduction}
\label{sec:SubbandDedispersion}
\textit{Sub-band dedispersion} is the name given to another technique used to
compute the dedispersion transform. Like the tree algorithm described in Section
\ref{sec:TreeDedispersionRoot}, the sub-band algorithm attempts to reduce the
cost of the computation relative to the direct method; however, rather than
exploiting a \textit{regularisation} of the dedispersion algorithm, the sub-band
method takes a simple \textit{approximation} approach.

In its simplest form, the algorithm involves two processing steps. In the first,
the set of trial DMs is approximated by a reduced set of $N_\textrm{DMnom} =
N_\textrm{DM}/N'_\textrm{DM}$ `nominal' DMs, each separated by $N'_\textrm{DM}$
trial dispersion measures. The \textit{direct} dedispersion algorithm is applied
to sub-bands of $N'_\nu$ channels to compute a dedispersed time series for each
nominal DM and sub-band. In the second step, the DM trials near each nominal
value are computed by applying the direct algorithm to the `miniature
filterbanks' formed by the time series for the sub-bands at each nominal
DM. These data have a reduced frequency resolution of $N_\textrm{SB} =
N_\nu/N'_\nu$ channels across the band. The two steps thus operate at reduced
dispersion measure and frequency resolution respectively, resulting in an
overall reduction in the computational cost.

The sub-band algorithm is implemented in the \textsc{presto} software suite
\citep{Ransom2001} and was recently implemented on a GPU by
\citet{MagroEtal2011} (see Section \ref{sec:other_work} for a comparison with
their work).
Unlike the tree algorithm, the sub-band method is able to compute the
dedispersion transform with the same flexibility as the direct method, making
its application to real observational data significantly simpler. 

The approximations made by the sub-band algorithm introduce additional smearing
into the dedispersed time series. We derive an analytic upper-bound in Appendix
\ref{sec:SubbandError} and show that, to first order, the smearing time
$t_\textrm{SB}$ is proportional to the product $N'_\textrm{DM} N'_\nu$ [see
  equation (\ref{eqn:subband_smear_approx})].

\subsection{Algorithm analysis}
\label{sec:subband_algo_analysis}
The computational complexity of the sub-band dedispersion algorithm can be
computed by summing that of the two steps:
\begin{align}
  T_\textrm{SB,1} &= N_\textrm{SB} \cdot
    T_\textrm{direct}(N_t, N'_\nu, N_\textrm{DMnom}) \\
  T_\textrm{SB,2} &= N_\textrm{DMnom} \cdot
    T_\textrm{direct}(N_t, N_\textrm{SB}, N'_\textrm{DM}) \\
  T_\textrm{SB} &= T_\textrm{SB,1} + T_\textrm{SB,2} \\
\label{eqn:subband_complexity}
    &= O \left[
      N_t N_\textrm{DM} N_\nu \left(
        \frac{1}{N'_\textrm{DM}} + \frac{1}{N'_\nu}
      \right)
    \right]
\end{align}
This result can be combined with knowledge of the smearing introduced by the
algorithm to probe the relationship between accuracy and performance. Inserting
the smearing constraint $t_\textrm{SB} \propto N'_\textrm{DM} N'_\nu$ (see
Section~\ref{sec:SubbandDedispersion}) into equation
(\ref{eqn:subband_complexity}), we obtain a second-order expression that is
minimised at $N'_\textrm{DM} = N'_\nu \propto \sqrt{t_\textrm{SB}}$, which
amounts to balancing the execution time between the two steps. This result
optimises the time complexity of the algorithm, which then takes the simple form
\begin{align}
\label{eqn:subband_complexity_optimal}
  T'_\textrm{SB} &= O \left( \frac{N_\textrm{DM} N_\nu}{\sqrt{t_\textrm{SB}}} \right)
\end{align}
and represents a theoretical speed-up over the direct algorithm proportional to
the square root of the introduced smearing.

The sub-band algorithm's dependence on the direct algorithm means that it
inherits similar algorithmic behaviour. However, as with the tree method, the
decrease in computational work afforded by the sub-band approach corresponds to
a decrease in the arithmetic intensity of the algorithm. This can be expected to
reduce the intrinsic performance of the two sub-band steps relative to the
direct algorithm.

One further consideration for the sub-band algorithm is the additional volume of
memory required to store the intermediate results produced by the first
step. These data consist of time series for each sub-band and nominal DM,
giving a space complexity of
\begin{align}
  M_\textrm{SB} = O \left( N_\textrm{SB} N_\textrm{DMnom} \right).
\end{align}
Assuming the time complexity is optimised as in equation
(\ref{eqn:subband_complexity_optimal}), the space complexity becomes
\begin{align}
  M'_\textrm{SB} = O \left( \frac{1}{t_\textrm{SB}} \right),
\end{align}
which indicates that the memory consumption increases much faster than the
execution time, in direct proportion to the introduced smearing rather than to
its square root. This can be expected to place a lower limit on the smearing
that can be achieved in practice.

\subsection{Implementation notes}
\label{sec:SubbandImplementation}
A significant advantage of the sub-band algorithm over the tree algorithm is
that it involves little more than repeated execution of the direct
algorithm. With sufficient generalisation\footnote{The direct dedispersion
  routine was modified to support `batching' (simultaneous application to
  several adjacent data-sets) and arbitrary strides through the input and output
  arrays, trial DMs and channels.} of our implementation of the direct
algorithm, we were able to implement the sub-band method with just two
consecutive calls to the direct dedispersion routine and the addition of a
temporary data buffer.

In our implementation, the `intermediate' data (i.e., the outputs of the first
step) are stored in the temporary buffer using 32 bits per sample. The second
call to the dedispersion routine then reads these values directly before writing
the final output using a desired number of bits per sample.

Experimentation showed that optimal performance occurred at a slightly different
shape and size of the thread blocks on the GPU compared to the direct algorithm
(see Section~\ref{sec:DirectAnalysis}). The sub-band kernels operated most
efficiently with 128 threads per block divided into 16 time samples and 8
DMs. In addition, the optimal choice of the ratio $N'_\nu/N'_\textrm{DM}$ was
found to be close to unity, which matches the theoretical result derived in
Section~\ref{sec:subband_algo_analysis}. While these parameters minimised the
execution time, the sub-band kernels were still found to perform around 40\%
less efficiently than the direct kernel. This result is likely due to the
reduced arithmetic intensity of the algorithm (see Section
\ref{sec:subband_algo_analysis}).

\section{Results}
\label{sec:Results}

\subsection{Smearing}
\label{sec:Accuracy}
\begin{figure*}
\centering
\subfigure[With time scrunching]{\includegraphics[width=8.5cm]{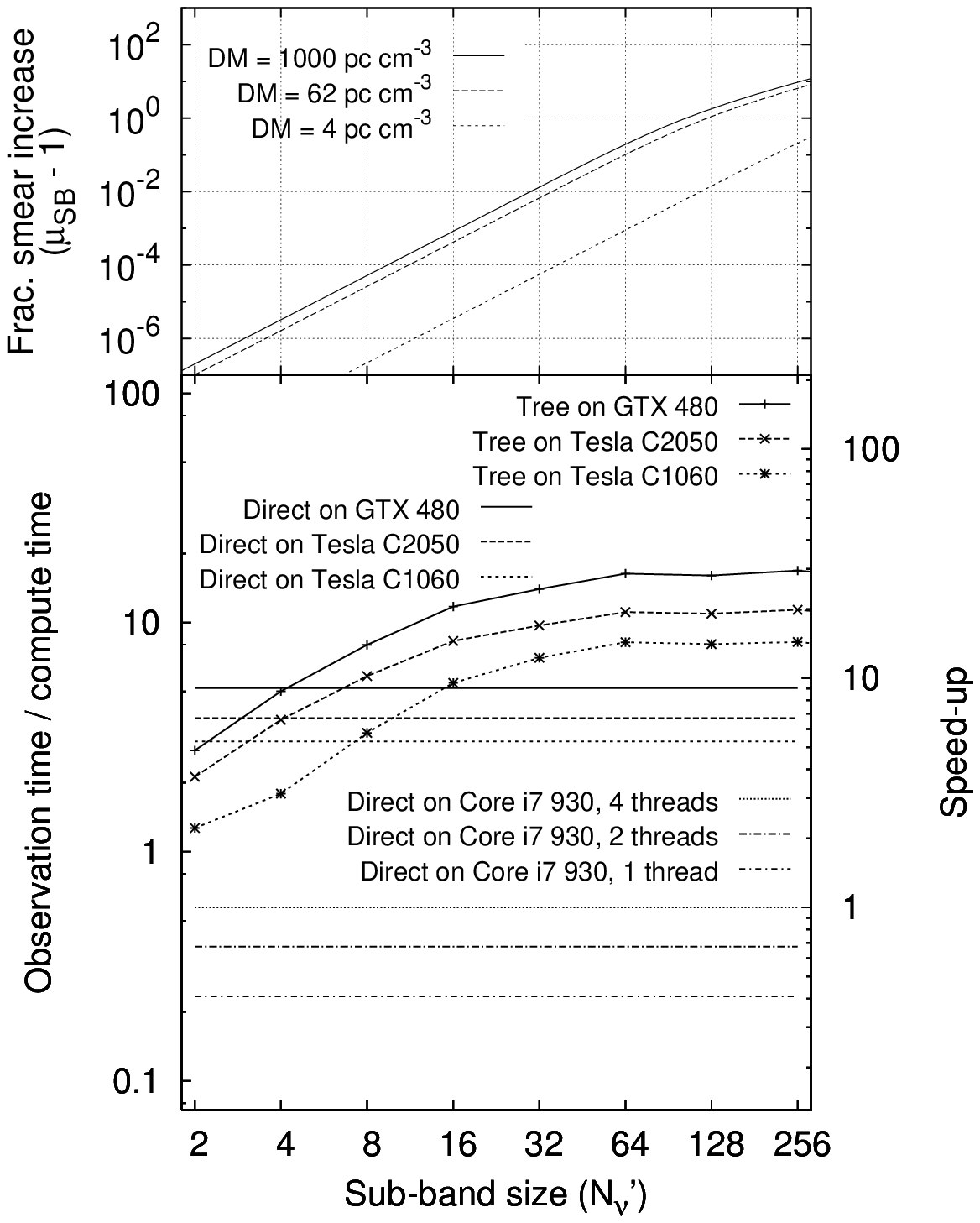}
                                 \label{fig:TreeResultsScrunch}}
\quad
\subfigure[Without time scrunching]{\includegraphics[width=8.5cm]{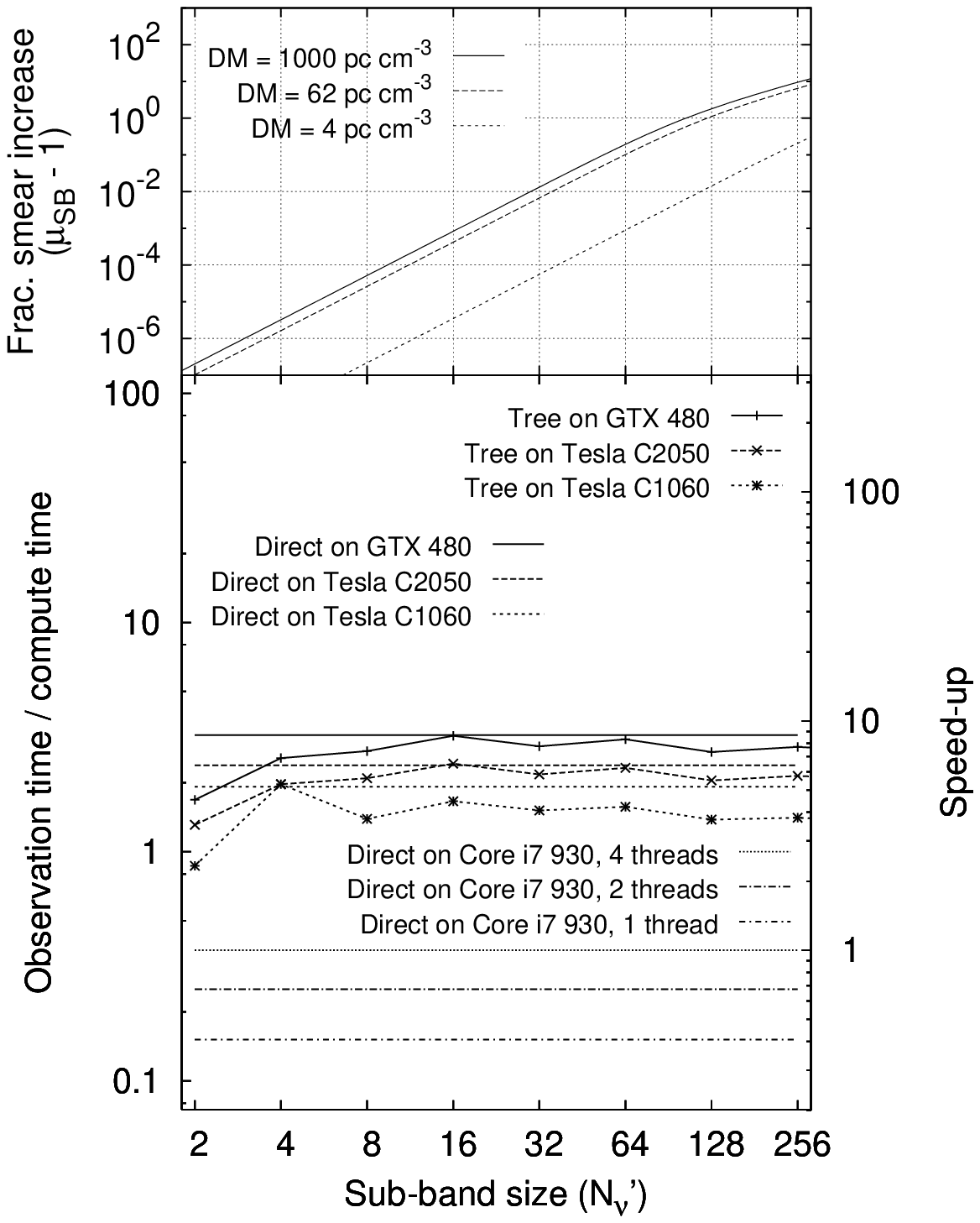}
                                 \label{fig:TreeResultsNoScrunch}}
\caption{ \textbf{Upper:} Analytic upper-bound on signal degradation of a
  40$\mu$s pulse due to the
  piecewise linear tree algorithm as a function of the number of channels per
  sub-band [see equation (\ref{eqn:RelativeSmearing})]. \textbf{Lower:}
  Performance results for the direct and piecewise linear tree algorithms with
  (a) and without (b) `time-scrunching' applied. Benchmarks were executed on an
  Intel Core i7 930 quad-core CPU and NVIDIA Tesla C1060, Tesla C2050 and
  GeForce GTX 480 GPUs. All results correspond to operations on one minute of
  input data with the following observing parameters: $\textrm{bits/sample}=2$,
  $\nu_c=1381.8\textrm{MHz}$, $\textrm{BW}=400\textrm{MHz}$, $N_\nu=1024$,
  $\Delta \tau=64 \mu\textrm{s}$.  A total of 1196 DM trials were used, spaced
  non-linearly in the range 0 $\le$ DM $<$ 1000 pc cm$^{-3}$ (see text for
  details). Error bars are too small to be seen at this scale and are not
  plotted. Note that performance results are projected from measurements of
  codes performing sub-sets of the benchmark task (see text for details).}
\label{fig:TreeResults}
\end{figure*}

\begin{figure*}
\centering
\subfigure[With time scrunching]{\includegraphics[width=8.5cm]{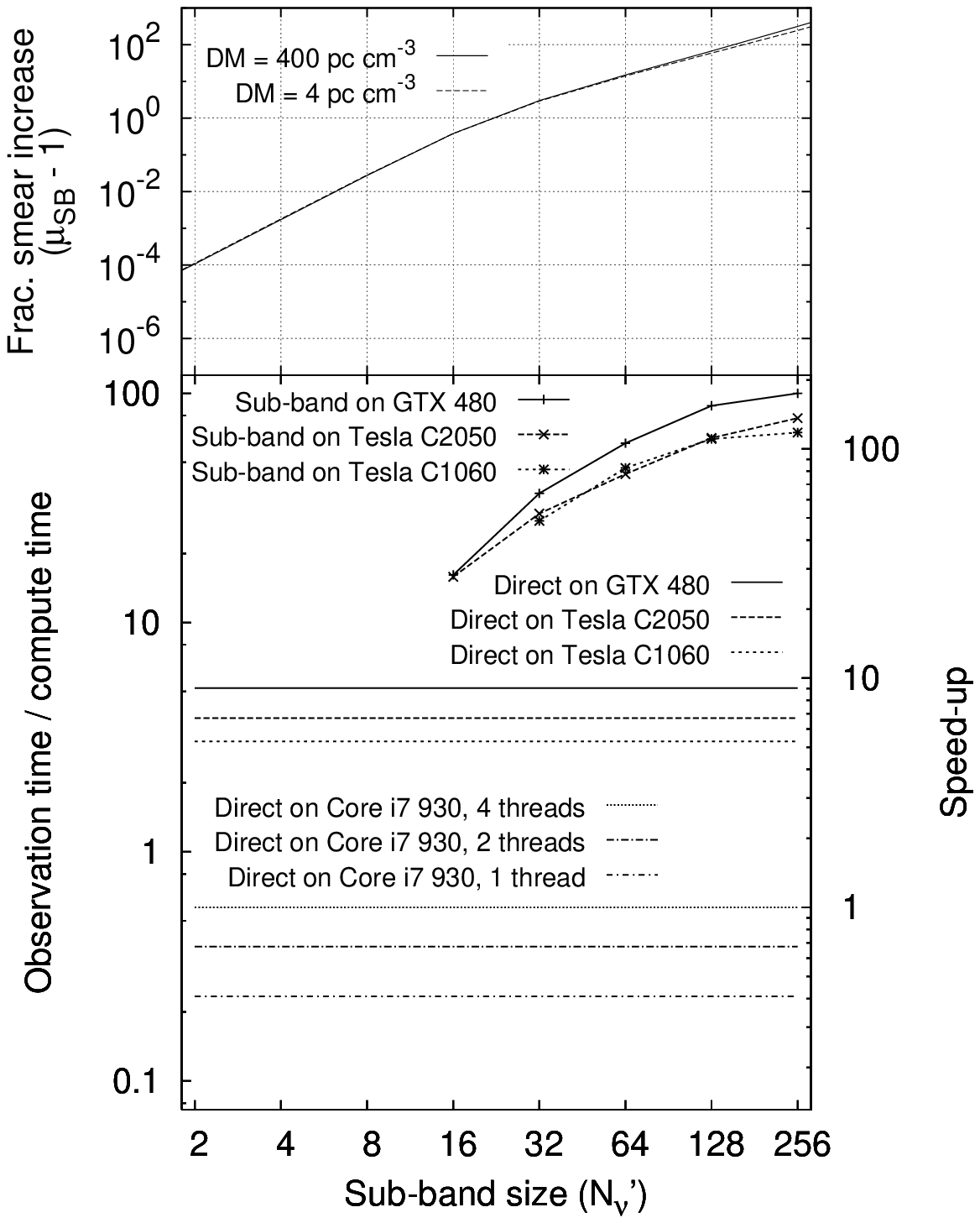}
                                 \label{fig:SubbandResultsScrunch}}
\quad
\subfigure[Without time scrunching]{\includegraphics[width=8.5cm]{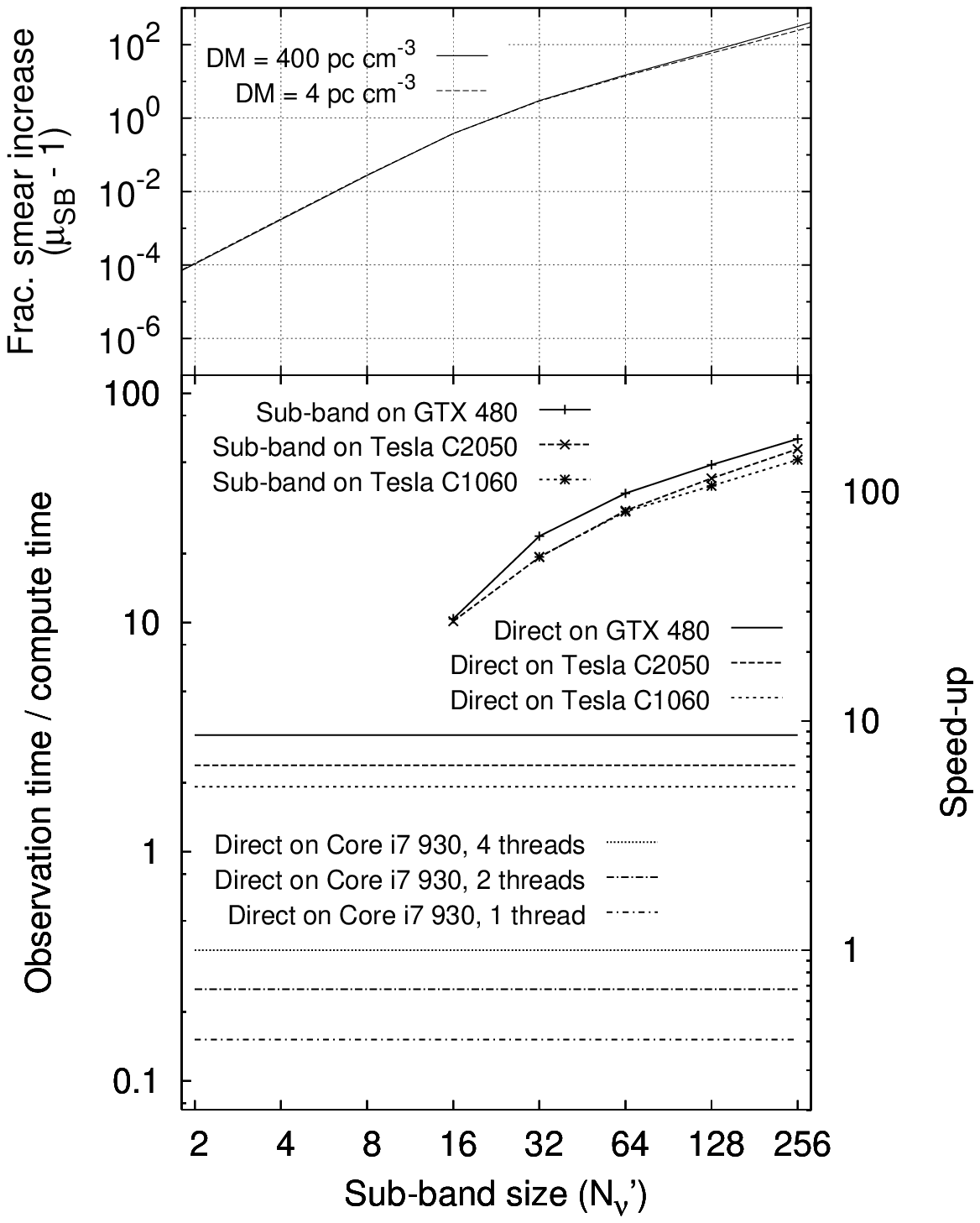}
                                    \label{fig:SubbandResultsNoScrunch}}
\caption{\textbf{Upper:} Analytic upper-bound on signal degradation of a
  40$\mu$s pulse due to the
  sub-band algorithm as a function of the number of channels per sub-band [see
    equation (\ref{eqn:RelativeSmearingSubband})]. \textbf{Lower:} Performance
  results for the direct and sub-band methods with (a) and without
  (b) the use of time-scrunching. See Fig.~\ref{fig:TreeResults}
  caption for details. Note that it was not possible to run benchmarks of the
  sub-band code for $N'_\nu < 16$ due to memory constraints.}
\label{fig:SubbandResults}
\end{figure*}

Our analytic upper-bounds on the increase in smearing due to use of the
piecewise linear tree algorithm [equation (\ref{eqn:RelativeSmearing})] and the
sub-band algorithm [equation (\ref{eqn:RelativeSmearingSubband})] are plotted in
the upper panels of Figs. \ref{fig:TreeResults} and \ref{fig:SubbandResults}
respectively. The reference point [$W$ in equations (\ref{eqn:RelativeSmearing})
  and (\ref{eqn:RelativeSmearingSubband})] was calculated using equations for
the smearing during the \textit{direct} dedispersion process\footnote{Levin,
  L. 2011, priv. comm.} assuming an intrinsic pulse width of 40$\mu$s.

For the piecewise linear tree algorithm, the effective signal smearing at low
dispersion measure is dominated by the intrinsic pulse width, the sampling time
$\Delta\tau$ and the effect of finite DM sampling. As the DM is increased,
however, the effects of finite channel width and the sub-band technique grow,
and eventually become dominant. These smearing terms both scale linearly with
the dispersion measure, and so the relative contribution of the sub-band method,
$\mu_\textrm{SB}$, tends to a constant.

The sub-band algorithm exhibits virtually constant smearing as a function of DM
due to its dependence on the DM step, which is itself chosen to maintain a fixed
fractional smearing. While the general trend mirrors that of the tree algorithm,
the sub-band algorithm's smearing is typically around two orders of magnitude
worse than its tree counterpart.

\subsection{Performance}
\label{sec:Performance}

Our codes as implemented allowed us to directly compute the following:
\begin{itemize}
  \item Any list of DMs using the direct or sub-band algorithm with no
    time-scrunching,
  \item DMs up to the diagonal [see equation (\ref{eqn:tree_dm_limit})] using
    the piecewise linear tree algorithm, and
  \item DMs up to the diagonal [see equation (\ref{eqn:padded_dm_diag})] using
  the frequency-padded tree algorithm.
\end{itemize}

A number of timing benchmarks were run to compare the performance of the CPU to
the GPU and the direct algorithm to the tree algorithms. Input and dispersion
parameters were chosen to reflect a typical scenario as appears in modern pulsar
surveys such as the High Time Resolution Universe (HTRU) survey currently
underway at the Parkes radio telescope \citep{KeithEtal2010}. The benchmarks
involved computing the dedispersion transform of one minute of input data with
observing parameters of $\textrm{bits/sample}=2$, $\nu_0=1581.8 \textrm{MHz}$,
$\Delta \nu=-0.39062 \textrm{MHz}$, $N_\nu=1024$, $\Delta \tau=64
\mu\textrm{s}$. DM trials were chosen to match those used in the HTRU survey,
which were originally derived by applying an analytic constraint on the
signal-smearing due to incorrect trial DM\footnote{Levin, L. 2011, priv. comm.;
  see \citealt{CordesMcLaughlin2003} for a similar derivation.}. The chosen set
contained 1196 trial DMs in the range 0 $\le$ DM $<$ 1000 pc cm$^{-3}$ with
approximately exponential spacing.

For comparison purposes, we benchmarked a reference CPU direct dedispersion code
in addition to our GPU codes. The CPU code (named \textsc{dedisperse\_all}) is
highly optimised, and uses multiple CPU cores to compute the dedispersion
transform (parallelised over the time dimension) in addition to bit-level
parallelism as described in Section~\ref{sec:DirectImplementation}.
\textsc{dedisperse\_all} is approximately 60$\times$ more efficient than the
generic \textsc{dedisperse} routine from
\textsc{sigproc}\footnote{sigproc.sourceforge.net}, but is only applicable to a
limited subset of data formats.

At the time of writing, our dedispersion code-base did not include
`full-capability' implementations of all of the discussed algorithms. However,
we were able to perform a number of benchmarks that were sufficient to obtain
accurate estimates of the performance of complete runs. Timing measurements for
our codes were projected to produce a number of derived results representative
of the complete benchmark task. The direct/sub-band dedispersion code was able
to compute the complete list of desired DMs, but was not able to exploit
time-scrunching; results for these algorithms with time scrunching were
calculated by assuming that the computation of DMs between 2$\times$ and
4$\times$ the diagonal would proceed twice as fast as the computation up to
2$\times$ the diagonal (as a result of there being half as many time samples),
and similarly for 4$\times$ to 8$\times$ etc. up to the maximum desired DM. A
simple code to perform the time-scrunching operation (i.e., adding adjacent time
samples to reduce the time resolution by a factor of two) was also benchmarked
and factored into the projection. For the tree codes, which were unable to
compute DMs beyond the diagonal, timing results were projected by scaling as
appropriate for the computation of the full set of desired DMs with or without
time-scrunching. Individual sections of code were timed separately to allow for
different scaling behaviours.

Benchmarks were run on a variety of hardware configurations. CPU benchmarks
were run on an Intel i7 930 quad-core CPU (Hyperthreading enabled). GPU
benchmarks were run using version 3.2 of the CUDA toolkit on the pre-Fermi
generation NVIDIA Tesla C1060 and the Fermi generation NVIDIA Tesla C2050
(error-correcting memory disabled) and GeForce GTX 480 GPUs.
Hardware specifications of the GPUs' host machines varied, but were not
considered to significantly impact performance measurements other than the
copies between host and GPU memory. Benchmarks for these copy operations
were averaged across the different machines.

Our derived performance results for the direct and piecewise linear tree codes
are plotted in the lower panels of Fig.~\ref{fig:TreeResults}. The performance
of the frequency-padded tree code corresponded to almost exactly half that of
the piecewise linear tree code at a sub-band size of $N'_\nu=1024$; these
results were omitted from the plot for clarity.

Performance results for the sub-band dedispersion code are plotted in the lower
panels of Fig.~\ref{fig:SubbandResults} along with the results of the direct
code for comparison. Due to limits on memory use (see Section
\ref{sec:subband_algo_analysis}), benchmarks for $N'_\nu < 16$ were not
possible.

Performance was measured by inserting calls to the Unix function
\textit{gettimeofday()} before and after relevant sections of code. Calls to the
CUDA function \textit{cudaThreadSynchronize()} were inserted where necessary to
ensure that asynchronous GPU functions had completed their execution prior to
recording the time.

Several different sections of code were timed independently. These included pre-
and post-processing steps (e.g., unpacking, transposing, scaling) and copies
between host and GPU memory (in both directions), as well as the dedispersion
kernels themselves.  Disk I/O and portions of code whose execution time does not
scale with the size of the input were not timed (see Section
\ref{sec:Discussion} for a discussion of the impact of disk I/O). Timing results
represent the total execution time of all timed sections, including memory
copies between the host and the device in the case of the GPU codes.

Each benchmark was run 101 times, from which the the median execution time was
chosen as the final measurement. Recorded uncertainties corresponded to the
5$^\textrm{th}$ and 95$^\textrm{th}$ percentiles; the error bars are too small
to be seen in Figs. \ref{fig:TreeResults} and \ref{fig:SubbandResults} and were
not plotted.

\section{Discussion}
\label{sec:Discussion}
The lower panel of Fig.~\ref{fig:TreeResultsScrunch} shows a number of
interesting performance trends. As expected, the slowest computation speeds come
from the direct dedispersion code running on the CPU. Here, some scaling is
achieved via the use of multiple cores, but the speed-up is limited to around
2.5$\times$ when using all four. This is likely due to saturation of the
available memory bandwidth.

Looking at the corresponding results on a GPU, a large performance advantage
is clear. The GTX 480 achieves a 9$\times$ speed-up over the quad-core CPU,
and even the last-generation Tesla C1060 manages a factor of 5$\times$. The
fact that a single GPU is able to compute the dedispersion transform in less
than a third of the real observation time makes it an attractive option for
real-time detection pipelines.

A further performance boost is seen in the transition to the tree
algorithm. Computation speed is projected to exceed that of the direct code
for almost all choices of $N'_\nu$, peaking at around 3$\times$ at
$N'_\nu=64$. Performance is seen to scale approximately linearly for
$N'_\nu<32$, before peaking and then decreasing very slowly for
$N'_\nu>64$. This behaviour is explained by the relative contributions of the
two stages of the computation. For small $N'_\nu$, the second, `sub-band
combination', stage dominates the total execution time [scaling as
  $O(1/N'_\nu)$]. At large $N'_\nu$ the execution time of the second stage
becomes small relative to the first, and scaling follows that of the basic
tree algorithm [i.e., $O(\log N'_\nu)$].

The results of the sub-band algorithm in Fig.~\ref{fig:SubbandResultsScrunch}
also show a significant performance advantage over the direct algorithm. The
computable benchmarks start at $N'_\nu$=16 with around the same performance as
the tree code. From there, performance rapidly increases as the size of the
sub-bands is increased, eventually tailing off around $N'_\nu$=256 with a
speed-up of approximately 20$\times$ over the direct code. At such high speeds,
the time spent in the GPU kernel is less than the time spent transferring the
data into and out of the GPU. The significance of this effect for each of the
three algorithms is given in Table \ref{tbl:CopyTimes}.

\begin{table}
\centering
\caption{Summary of host$\leftrightarrow$GPU memory copy times}
\begin{tabular}{lrr}
\label{tbl:CopyTimes}
Code     & Copy Time & Fraction of total time \\
\hline
Direct   & 0.62 s          & $< 5\%$ \\
Tree     & 1.05 s          & $< 30\%$ \\
Sub-band & 0.62 s          & 10\% -- 65\%
\end{tabular}
\end{table}

The results discussed so far have assumed the use of the time-scrunching
technique during the dedispersion computation. If time-scrunching is
\textit{not} used, the projected performance results change significantly [see
  lower panels Figs. \ref{fig:TreeResultsNoScrunch} and
  \ref{fig:SubbandResultsNoScrunch}]. Without the use of time-scrunching, the
direct dedispersion codes perform around 1.6$\times$ slower, and similar results
are seen for the sub-band code. The tree codes, however, are much more severely
affected, and perform 5$\times$ slower when time-scrunching is not
employed. This striking result can be explained by the inflexibilities of the
tree algorithm discussed in Section~\ref{sec:TreeDedispersion}. At large
dispersion measure, the direct algorithm allows one to sample DM space very
thinly. The tree algorithms, however, do not -- they will always compute DM
trials at a fixed spacing [see equation (\ref{eqn:tree_dms})]. This means that
the tree algorithms are effectively over-computing the problem, which leads to
the erosion of their original theoretical performance advantage. The use of
time-scrunching emulates the thin DM-space sampling of the direct code, and
allows the tree codes to maintain an advantage.

While the piecewise linear tree code and the sub-band code are seen to provide
significant speed-ups over the direct code, their performance leads come at the
cost of introducing additional smearing into the dedispersed signal. Our
analytic results for the magnitude of the smearing due to the tree code (upper
panels Fig.~\ref{fig:TreeResults}) show that for the chosen observing
parameters, the total smear is expected to increase by less than 10\% for all
$N'_\nu \leq 64$ at a DM of 1000 pc cm$^{-3}$. Given that peak performance of
the tree code also corresponded to $N'_\nu = 64$, we conclude that this is the
optimal choice of sub-band size for such observations.

The smearing introduced by the sub-band code (upper panels
Fig.~\ref{fig:SubbandResults}) is significantly worse, increasing the signal
degradation by three orders of magnitude more than the tree code. Here, the
total smear is expected to increase by around 40\% at $N'_\nu$=16, and at
$N'_\nu$=32 the increase in smearing reaches 300\%. While these results are
upper limits, it is unlikely that sub-band sizes of more than $N'_\nu$=32 will
produce acceptable results in practical scenarios.

In contrast to the piecewise linear code, the frequency-padded tree code showed
only a modest speed-up of around 1.5$\times$ over the direct approach due to its
doubling of the number of frequency channels. Given that the sub-band algorithm
has a minimal impact on signal quality, we conclude that the frequency-padding
technique is an inferior option.

It is also important to consider the \textit{development cost} of the algorithms
we have discussed. While the tree code has shown both high performance and
accuracy, it is also considerably more complex than the other algorithms. The
tree algorithm in its base form, as discussed in
Section~\ref{sec:TreeDedispersion}, is much less intuitive than the direct
algorithm (e.g., the memory access patterns in
Fig.~\ref{fig:TreeAlgorithm}). This fact alone makes implementation more
difficult. The situation gets significantly worse when one must adapt the tree
algorithm to work in practical scenarios, with quadratic dispersion curves and
arbitrary DM trials. Here, the algorithm's inflexibility makes implementation a
daunting task. We note that our own implementations are as yet incomplete. By
comparison, implementation of the direct code is relatively straightforward, and
the sub-band code requires only minimal changes. Development time must play a
role in any decision to use one dedispersion algorithm over another.

The three algorithms we have discussed each show relative strengths and
weaknesses. The direct algorithm makes for a relatively straightforward move to
the GPU architecture with no concerns regarding accuracy, and offers a speed-up
of up to 10$\times$ over an efficient CPU code. However, its performance is
convincingly beaten by the tree and sub-band methods. The tree method is able to
provide significantly better performance with only a minimal loss of signal
quality; however, it comes with a high cost of development that may outweigh its
advantages. Finally, the sub-band method combines excellent performance with an
easy implementation, but is let down by the substantial smearing it introduces
into the dedispersed signal. The optimal choice of algorithm will therefore
depend on which factors are most important to a particular project. While there
is no clear best choice among the three different algorithms, we emphasize that
between the two hardware architectures the GPU clearly outperforms the CPU.

When comparing the use of a GPU to a CPU, it is interesting to note that our
final GPU implementation of the direct dedispersion algorithm on a Fermi-class
device is, relatively speaking, a simple code. While it was necessary in both
the pre-Fermi GPU and multi-core CPU implementations to use non-trivial
optimisation techniques (e.g., texture memory, bit-packing etc.), the optimal
implementation on current-generation, Fermi, GPU hardware was also the simplest
or `obvious' implementation. This demonstrates how far the (now rather misnamed)
graphics processing unit has come in its ability to act as a general-purpose
processor.

In addition to the performance advantage offered by GPUs today, we expect our
implementations of the dedispersion problem to scale well to future
architectures with little to no code modification. The introduction of the
current generation of GPU hardware brought with it both a significant
performance increase and an equally significant reduction in programming
complexity. We expect these trends to continue when the next generation of GPUs
is released, and see a promising future for these architectures and the
applications that make use of them.

While we have only discussed single-GPU implementations of dedispersion, it
would in theory be a simple matter to make use of multiple GPUs, e.g., via
time-division multiplexing of the input data or allocation of a sub-set of beams
to each GPU. As long as the total execution time is dominated by the GPU
dedispersion kernel, the effects of multiple GPUs within a machine sharing
resources such as CPU cycles and PCI-Express bandwidth are expected to be
negligible. However, as shown in Table~\ref{tbl:CopyTimes}, the tree and
sub-band codes are in some circumstances so efficient that
host$\leftrightarrow$device memory copy times become a significant fraction of
the total run time. In these situations, the use of multiple GPUs within a
single host machine may influence the overall performance due to reduced
PCI-Express bandwidth.

Disk I/O is another factor that can contribute to the total execution time of a
dedispersion process. Typical server-class machines have disk read/write speeds
of only around 100 MB/s, while our GPU dedispersion codes are capable of
producing 8-bit time series at well over twice this rate. If dedispersion is
performed in an offline fashion, where time series are read from and written to
disk before and after dedispersion, then it is likely that disk performance will
become the bottle-neck. The use of multiple GPUs within a machine may
exacerbate this effect. However, for real-time processing pipelines where data
are kept in memory between operations, the dedispersion kernel can be expected
to dominate the execution time. This is particularly important for transient
search pipelines, where acceleration searching is not necessary and dedispersion
is typically the most time-consuming operation.

The potential impact of limited PCI-Express bandwidth or disk I/O performance
highlights the need to remember Amdahl's Law when considering further speed-ups
in the dedispersion codes: the achievable speed-up is limited by the largest
bottle-neck. The tree and sub-band codes are already on the verge
of being dominated by the host$\leftrightarrow$device memory copies, meaning
that further optimisation of their kernels will provide diminishing
returns. While disk and memory bandwidths will no-doubt continue to increase in
the future, we expect the ratio of arithmetic performance to memory performance
to get worse rather than better.

The application of GPUs to the problem of dedispersion has produced
speed-ups of an order of magnitude. The implications of this result for current
and future surveys are significant. Current projects often execute pulsar and
transient search pipelines offline due to limited computational resources. This
results in event detections being made long after the time of the events
themselves, limiting analysis and confirmation power to what can be gleamed from
archived data alone. A real-time detection pipeline, made possible by a
GPU-powered dedispersion code, could instead trigger systems to record
invaluable baseband data during significant events, or alert other observatories
to perform follow-up observations over a range of wavelengths.  Real-time
detection capabilities will also be crucial for next-generation telescopes such
as the Square Kilometre Array pathfinder programs ASKAP and MeerKAT. The use of
GPUs promises significant reductions in the set-up and running costs of
real-time pulsar and transient processing pipelines, and could be the enabling
factor in the construction of ever-larger systems in the future.

\subsection{Comparison with other work}
\label{sec:other_work}
\citet{MagroEtal2011} recently reported on a GPU code that could achieve very
high ($>100\times$) speed-ups over the dedispersion routines in \textsc{sigproc}
and \textsc{presto} \citep{Ransom2001} whereas our work only finds improvements
of factors of 10--30 over \textsc{dedisperse\_all}. There are two key reasons
for the apparent discrepancy in speed. Firstly, the \textsc{sigproc} routine
was never written to optimise performance but rather to produce reliable
dedispersed data streams from a very large number of different
backends. Inspection of the innermost loop reveals a conditional test that
prohibits parallelisation, and a two dimensional array that is computationally
expensive. Secondly, \textsc{sigproc} only produces one DM per file read, which
is very inefficient. We believe that these factors explain the very large
speed-ups reported by Magro et al. In our own benchmarks, we have found our CPU
comparison code \textsc{dedisperse\_all} to be $\sim 60\times$ faster than
\textsc{sigproc}. For comparison, this puts our direct GPU code at $\sim
300\times$ faster than \textsc{sigproc} when using the same Tesla C1060 model
GPU as Magro et al.

Direct comparison of our GPU results with those of Magro et al. is difficult, as
the details of the CPU code, the method of counting FLOP/s and the observing
parameters used in their performance plots is not clear. However, we have
benchmarked our GPU code on the `toy observation' presented in section 5 of
their paper. The execution times are compared in
Table~\ref{tbl:ToyObservation}. Magro et al. did not specify the number of bits
per sample used in their benchmark; we chose to use 8 bits/sample, but found no
significant difference when using 32 bits/sample. We found our implementation of
the direct dedispersion algorithm to be $\sim 2.3\times$ faster than that
reported in their work. Possible factors contributing to this difference
include our use of texture memory, two-dimensional thread blocks and allocation
of multiple samples per thread.
The performance results of our implementation of the sub-band dedispersion
algorithm generally agree with those of Magro et al., although the impact of the
additional smearing is not quantified in their work.

\begin{table}
\centering
\caption{Timing comparisons for direct GPU dedispersion of the `toy
  observation' defined in \citet{MagroEtal2011} ($\nu_c$=610 MHz, BW=20 MHz,
  $N_\nu$=256, $\Delta\tau$=12.8$\mu$s, $N_\textrm{DM}$=500, 0$\le$DM$<$60 pc
  cm$^{-3}$). All benchmarks were executed on a Tesla C1060 GPU.}
\begin{tabular}{lrrr}
\label{tbl:ToyObservation}
Stage                    & \citet{MagroEtal2011} & This work        & Ratio \\
\hline
Corner turn              & 112 ms                & 7 ms             & 16$\times$ \\
\textbf{De-dispersion}   & \textbf{4500 ms}      & \textbf{1959 ms} & \textbf{2.29$\times$} \\
GPU$\rightarrow$CPU copy & 220 ms                & 144 ms           & 1.52$\times$ \\
\hline
Total                    & 4832 ms               & 2110 ms          & 2.29$\times$
\end{tabular}
\end{table}

In summary, we agree with Magro et al. that GPUs offer great promise in
incoherent dedispersion. The benefit over that of CPUs is, however, closer to
the ratio of their memory bandwidths ($\sim 10 \times$) than the factor of 100
reported in their paper, which relied on comparison with a non-optimised
single-threaded CPU code.

\subsection{Code availability}
We have packaged our GPU implementation of the direct incoherent dedispersion
algorithm into a C library that we make available to the
community\footnote{Our library and its source code are available at:
\url{http://dedisp.googlecode.com/}}. The application
programming interface (API) was modeled on that of the FFTW
library\footnote{\url{http://www.fftw.org}}, which was found to be a
convenient fit. The library requires the NVIDIA CUDA Toolkit, but places no
requirements on the host application, allowing easy integration into
existing C/C++/Fortran etc. codes. While the library currently uses the
direct dedispersion algorithm, we may consider adding support for a tree
or sub-band algorithm in future.

\section{Conclusions}
\label{sec:Conclusion}
We have analysed the direct, tree and sub-band dedispersion algorithms and found
all three to be good matches for massively-parallel computing architectures such
as GPUs. Implementations of the three algorithms were written for the current
and previous generations of GPU hardware, with the more recent devices providing
benefits in terms of both performance and ease of development. Timing results
showed a 9$\times$ speed-up over a multi-core CPU when executing the direct
dedispersion algorithm on a GPU. Using the tree algorithm with a piecewise
linear approximation technique results in some additional smearing of the input
signal, but was projected to provide a further 3$\times$ speed-up at a very
modest level of signal-loss. The sub-band method provides a means of obtaining
even greater speed-ups, but imposes significant additional smearing on the
dedispersed signal. These results have significant implications for current and
future radio pulsar and transient surveys, and promise to dramatically lower the
cost barrier to the deployment of real-time detection pipelines.

\section*{Acknowledgments}
We would like to thank Lina Levin and Willem Van Straten
for very helpful discussions relating to pulsar searching, Mike Keith for
valuable information regarding the tree dedispersion algorithm, and Paul
Coster for help in testing our dedispersion code.
We would also like to thank the referee Scott Ransom for his very helpful
comments and suggestions for the paper.

\appendix

\section{Error analysis for the tree dedispersion algorithm}
\label{sec:TreeError}
Here we derive an expression for the maximum error introduced by the use of
the piecewise linear tree dedispersion algorithm.

The deviation of a function $f(x)$ from a linear approximation between $x=x_0$
and $x=x_1$ is bounded by
\begin{align}
\label{eqn:linear_interp}
  \varepsilon_f &\le \frac{1}{8}(x_1 - x_0)^2 \max_{x_0\le x\le x_1} \left|
  \frac{d^2}{dx^2}f(x) \right|,
\end{align}
which shows that the error is proportional to the square of the step size and
the second derivative of the function. For the dedispersion problem, the second
derivative of the delay function with respect to frequency is given by
\begin{align}
\label{eqn:delay_derivative}
  \frac{\partial^2}{\partial\nu^2}\Delta t(d,\nu) &= \textrm{DM}(d)\frac{d^2}{d\nu^2}\Delta
  T(\nu) \\
  &= 6\; \textrm{DM}(d)
  k_\textrm{DM} \frac{\Delta\nu^2}{\nu_0^4}
  \left( 1+\frac{\Delta\nu}{\nu_0}\nu \right)^{-4},
\end{align}
which has greater magnitude at lower frequencies. Evaluating at the lowest
frequency in the band, $\nu=N_\nu$, and substituting into equation
(\ref{eqn:linear_interp}) along with the sub-band size $N'_\nu$, one finds the
error to be bounded by:
\begin{align}
\label{eqn:subband_error}
  t_\textrm{tree} \equiv \varepsilon_{\Delta t} \le \frac{3}{4} \textrm{DM}
  \frac{k_\textrm{DM}}{\nu_0^2} \left( \frac{N'_\nu}{N_\nu} \right)^2
  \frac{\lambda^2}{(1+\lambda)^4},
\end{align}
where $\lambda \equiv \frac{\Delta\nu}{\nu_0}N_\nu$ is a proxy for the
\textit{fractional bandwidth}, a measure of the width of the antenna band.

If the smearing as a result of using the direct algorithm is quantified as the
effective width, $W$, of an observed pulse, then the piecewise linear tree
algorithm is expected to produce a signal with an effective width of
\begin{align}
  W_\textrm{tree} = \sqrt{W^2 + t^2_\textrm{tree}},
\end{align}
giving a relative smearing of
\begin{align}
\label{eqn:RelativeSmearing}
  \mu_\textrm{tree} \equiv \frac{W_\textrm{tree}}{W} = \frac{\sqrt{W^2 +
      t^2_\textrm{tree}}}{W}.
\end{align}
In contrast to the use of a piecewise linear approximation, the use of a change
of frequency coordinates (`frequency padding') to linearise the dispersion
trails results in no additional sources of smear.

\section{Error analysis for the sub-band dedispersion algorithm}
\label{sec:SubbandError}
Here we derive an expression for the maximum error introduced by the
use of the sub-band dedispersion algorithm.

The smearing introduced into a dedispersed time series due to an
approximation to the dispersion curve is bounded by the maximum
temporal deviation of the approximation from the exact curve. The
maximum change in delay across a sub-band is $\Delta t(\textrm{DM},
N_\nu) - \Delta t(\textrm{DM}, N_\nu-N'_\nu)$; the difference in this
value between two nominal DMs then gives the smearing time:
\begin{align}
  t_\textrm{SB} &\le 
  \Delta \textrm{DM}_\textrm{nom}
    \left[ \Delta T(N_\nu) - \Delta T(N_\nu-N'_\nu)\right] \\
\label{eqn:subband_smear_approx}
  &= N'_\textrm{DM} \Delta \textrm{DM} \frac{k_\textrm{DM}}{\nu_0^2}
    \left[ -2\frac{N'_\nu}{N_\nu} \frac{\lambda}{(1+\lambda)^3}
           + O \left( \frac{N'_\nu}{N_\nu} \right)^2
    \right ],
\end{align}
where the second form is obtained through Taylor expansion in powers of
$\frac{N'_\nu}{N_\nu}$ around zero.
Note that this derivation assumes the dispersion curve is approximated
by aligning the 'early' edge of each sub-band. An alternative approach is
to centre the sub-bands on the curve, which reduces the smearing by
$\sim 2\times$ but adds complexity to the implementation.

As with the tree algorithm, we can define the relative smearing of the sub-band
algorithm with respect to the direct algorithm as 
\begin{align}
\label{eqn:RelativeSmearingSubband}
  \mu_\textrm{SB} \equiv \frac{W_\textrm{SB}}{W} = \frac{\sqrt{W^2 +
      t^2_\textrm{SB}}}{W},
\end{align}
where, as before, $W$ is the effective width of an observed pulse after direct
dedispersion.

\bibliography{abbrevs,benbarsdell}{}
\bibliographystyle{mn2e_modded}

\label{lastpage}

\end{document}